\pdfoutput=1

\documentclass[11pt]{article}

\usepackage{tcolorbox}
\usepackage{amssymb}
\usepackage[preprint]{acl}

\newcommand{\xhdr}[1]{\vspace{0.3em}\noindent{{\bf #1.}}}
\usepackage{times}
\usepackage{latexsym}
\usepackage{bbding}
\usepackage[T1]{fontenc}
\usepackage{booktabs}
\usepackage[utf8]{inputenc}
\usepackage{xcolor}
\usepackage{microtype}
\usepackage{colortbl}

\usepackage{inconsolata}
\usepackage{multirow}
\usepackage{graphicx}

\title{Video2Roleplay: A Multimodal Dataset and Framework for Video-Guided Role-playing Agents}

\author{Xueqiao Zhang, Chao Zhang, Jingtao Xu, Yifan Zhu, Xin Shi, Yi Yang, Yawei Luo\thanks{\ \ Corresponding author}\\
  Zhejiang University \\
  \texttt{\{xueqiaozhang, yaweiluo\}@zju.edu.cn}}

\begin{document}
\maketitle
\begin{abstract}
Role-playing agents (RPAs) have attracted growing interest for their ability to simulate immersive and interactive characters. However, existing approaches primarily focus on static role profiles, overlooking the dynamic perceptual abilities inherent to humans. To bridge this gap, we introduce the concept of dynamic role profiles by incorporating video modality into RPAs. To support this, we construct \textbf{Role-playing-Video60k}, a large-scale, high-quality dataset comprising 60k videos and 700k corresponding dialogues. Based on this dataset, we develop a comprehensive RPA framework that combines adaptive temporal sampling with both dynamic and static role profile representations. Specifically, the dynamic profile is created by adaptively sampling video frames and feeding them to the LLM in temporal order, while the static profile consists of (1) character dialogues from training videos during fine-tuning, and (2) a summary context from the input video during inference. This joint integration enables RPAs to generate greater responses. Furthermore, we propose a robust evaluation method covering eight metrics. Experimental results demonstrate the effectiveness of our framework, highlighting the importance of dynamic role profiles in developing RPAs.\footnote{Our data and code are available at \url{https://github.com/zxqSled/Video2Roleplay}.}
\end{abstract}

\begin{figure}[t]
  \centering \includegraphics[width=\linewidth]{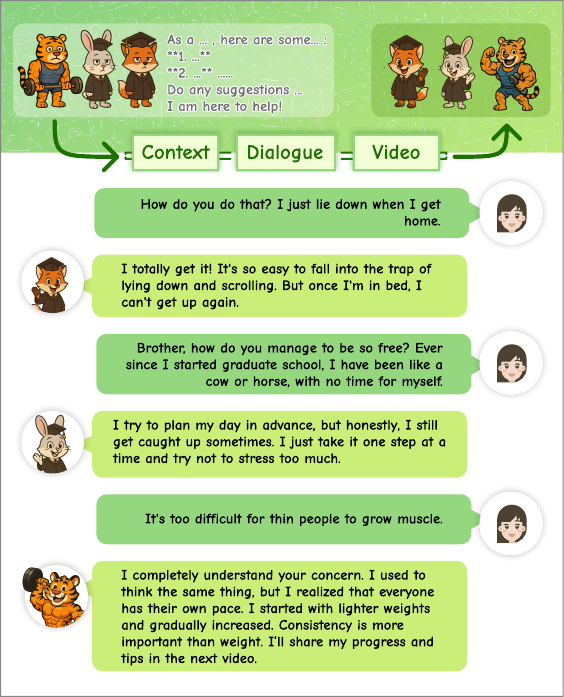}
  \caption{Examples illustrating our RPAs' performance compared to general baselines. More examples are provided in Appendix.}
  \label{fig:data}
  \vspace{-15pt}
\end{figure}
\section{Introduction}

Recent advancements in large language models (LLMs)~\cite{llmsurvey} have spurred significant research interest in RPAs~\cite{rolesurvey}, which simulate interactive characters through the integration of diverse modality data to create realistic user experiences.
However, real-world human perception is inherently multifaceted and dynamic. The current reliance primarily on static modalities like text and images limits the ability of these agents to fully satisfy the growing demand for highly immersive and expressive role-playing experiences.

Video, as a powerful multimodal medium~\cite{videoemotion,yang2024videoscene,lian2024videollm,mou2024revideomotion}, offers a rich array of dynamic details related to characters, such as emotional states, physical actions, scene transitions, and narrative experiences. This information is highly valuable for pioneering dynamic role-playing profiles.
For example, lives showcase character dynamic motions in authentic scenarios. Vlogs and role documentaries capture individuals' expressions and daily activities, effectively conveying complex emotions and personality traits for detailed character portrayals. Consequently, integrating the video modality into RPAs equips agents with more comprehensive and detailed dynamic information, improving role-playing performance and user engagement.

Currently, despite some promising results of the existing work~\cite{dai2024mmrole,wang2025coser} in the field of RPAs, there is still a lack of exploration in data resources and effective methods of video modality. How to effectively integrate video modality information with existing static modalities and leverage its unique dynamic information advantages for RPAs remains a challenging problem. Furthermore, the long length of some videos often introduces considerable redundant information, leading to high memory resource consumption and inefficient video information representation.

To fill these gaps, this study introduces the concept of dynamic role-playing to integrate video modality into the RPAs for the first time, constructs a large-scale and high-quality dataset tailored to the requirements of dynamic profile representation in RPAs, and proposes a comprehensive framework that effectively incorporates video modality with static modalities. 

Specifically, we construct a large-scale and high-quality dataset sourced from various social media platforms like Xiaohongshu, Douyin, Weibo, and Bilibili. The dataset comprises daily lives, lifestyle vlogs, and personal documentaries from diverse groups, accompanied by corresponding video captions and related dialogues, providing rich resources for the development of RPAs. Additionally, we propose a novel multimodal RPA framework that combines adaptive temporal sampling with both dynamic and static role profile representations. To construct the dynamic role profile, we adaptively sample video frames based on their duration and provide them to the LLM in their original order. In parallel, the static role profile captures character information with two main components: (1) character-specific dialogues related to training videos, which are used to guide the base model during fine-tuning, and (2) a high-level summary generated from the input video during inference, which provides a concise but accurate description of the video scene and character presentation. By integrating both dynamic and static role profiles, our framework enables RPAs to generate responses that are highly consistent with the character’s identity and the narrative context.

Moreover, we design a series of evaluation metrics and experiments to validate the effectiveness of our framework. Extensive experiments demonstrate the superior performance of our framework on RPAs. It establishes a compelling trade-off between parameter size and overall performance while achieving SOTA for human-likeness.

In summary, our contributions are threefold:
\begin{itemize}
\item{We are the first to integrate the video modality into RPAs, introducing the concept of dynamic role-playing and enabling the creation of rich dynamic role profiles.}
\item{We construct a large-scale and high-quality dataset for the development of RPAs, including $60k$ videos and $700k$ dialogues across various categories, durations, and scenarios.}
\item{We develop a novel and comprehensive RPA framework that integrates adaptive temporal sampling with both dynamic and static role profiles. Extensive experiments and analyses demonstrate its outstanding performance.}
\end{itemize}

\begin{table}
    \centering
    \caption{\textbf{Comparison between different role-playing datasets.} Our work is the first role-playing dataset that introduces the video. 
    \label{tab:dataset}} 
     \resizebox{\columnwidth}{!}{%
	\begin{tabular}{ccccc}
	\toprule
        Dataset  & Dialogues  & Video   \\
        \midrule
        ChatHaruhi~\cite{li2023chatharuhi} & 54,726 &     \textcolor{red}{\XSolidBrush}   \\
        Character-LLM~\cite{characterllm} &  14,300 &     \textcolor{red}{\XSolidBrush}   \\
		RoleLLM~\cite{wang2024rolellm} &  168.1k  & \textcolor{red}{\XSolidBrush} \\
        CharacterGLM~\cite{zhou2024characterglm} &  1,034 &   \textcolor{red}{\XSolidBrush}   \\
        Character100~\cite{character100} &  10,609   & \textcolor{red}{\XSolidBrush}   \\
        DITTO~\cite{lu2024ditto} &  7,186  & \textcolor{red}{\XSolidBrush}   \\
        CharacterEval~\cite{tu2024charactereval} &   1785    & \textcolor{red}{\XSolidBrush}  \\
        LifeChoice~\cite{lifechoice} &  1,462   & \textcolor{red}{\XSolidBrush}  \\
        RolePersonality~\cite{ran2024rolepersonality} &  87,345   & \textcolor{red}{\textcolor{red}{\XSolidBrush} }  \\ 
        MMRole~\cite{dai2024mmrole} &  14,346   & \textcolor{red}{\XSolidBrush}   \\           
        CharacterBench~\cite{zhou2025characterbench}  &  13,162   & \textcolor{red}{\XSolidBrush} \\  
        OpenCharacter~\cite{wang2025opencharacter}  & 306k    & \textcolor{red}{\XSolidBrush}   \\ 
        RoleMRC~\cite{LUandLI2025RoleMRC}  & 39.3k   & \textcolor{red}{\XSolidBrush}   \\
        CoSER~\cite{wang2025coser} &  29,798  &  \textcolor{red}{\XSolidBrush}    \\ 
        \midrule
		\rowcolor{blue!5}\textbf{Role-playing-Video60k(Ours)} &  700k & \textcolor{green}{\CheckmarkBold} \\
		\bottomrule
	\end{tabular}
    }
\end{table}
\section{Related Work}
\subsection{Static Role Playing}
ChatHaruhi~\cite{li2023chatharuhi} provides a dataset of over $54k$ simulated dialogues for $32$ characters spanning Chinese, English, and anime.
CharacterGLM~\cite{zhou2024characterglm} allows for personalizing a diverse range of agent personas and social agents through customizable attributes and behaviors.
CharacterLLM~\cite{characterllm} builds a dataset detailing specific character experiences, then fine-tunes a base model with the dataset to achieve target character portrayal.
RoleLLM~\cite{wang2024rolellm} improves LLM role-playing via a multi-component framework (e.g., role profile construction, role-GPT, role-bench).
Ditto~\cite{lu2024ditto} introduces a self-alignment method to enhance LLM role-playing capabilities through knowledge augmentation and dialogue simulation.
MMrole~\cite{dai2024mmrole}introduces the concept of multimodal role-playing agents and offers a comprehensive framework for their development and evaluation.
RoleMRC~\cite{LUandLI2025RoleMRC} provides a fine-grained composite benchmark for role-playing and instruction-following, revealing activation patterns linked to these distinct abilities.
CoSER~\cite{wang2025coser} provides a dataset comprising $29,798$ authentic conversations and comprehensive data from 771 renowned books and proposes a given-circumstance acting method for training and evaluating role-playing LLMs.

\subsection{Video Understanding}
GPT4Video~\cite{wang2024gpt4video} proposes a unified framework for video understanding and generation via pre-trained model integration and develops a simple text-only fine-tuning method for instruction following and safety alignment.
LongVLM~\cite{weng2024longvlm} introduces a VideoLLM for long-term video understanding, achieving affordability via segment decomposition, feature extraction, token merging, and global semantics.
Video-LLaVA~\cite{videollava} maps visual signals to the language feature space to achieve unified visual representations, introducing a method for aligning features prior to projection.
VideoAgent~\cite{wang2024videoagent} proposes an agent-based system that iteratively extracts and compiles key information for question answering, using vision-language models for visual translation and retrieval.
VidRecap~\cite{Recap} proposes a hierarchical caption generation method that creates CLIP captions, segment descriptions, and video summaries, trained using a coarse-to-fine approach to learn the structure of video.
LongVU~\cite{shen2024longvu} preserves frame information for lengthy videos by compressing tokens based on similarity and selecting relevant visual tokens for text queries.
InternVideo2.5~\cite{wang2025internvideo2.5} introduces a length-adaptive token approach to process videos, integrating visual perception with MLLM for fine-grained analysis.
\begin{figure}[t]
  \centering \includegraphics[width=\linewidth]{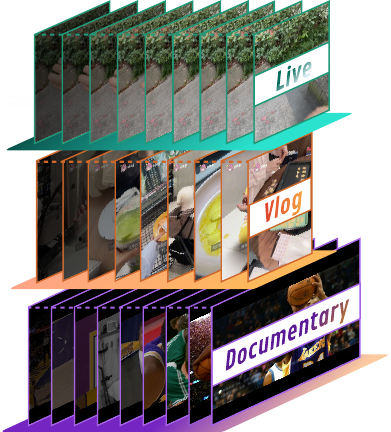}
  \caption{The video types and examples of our dataset.}
  \vspace{-10pt}
  \label{fig:type}
\end{figure}
\subsection{Multimodal Large Language Model}
CLIP~\cite{CLIP} achieves cross-modal understanding and unified representation by applying contrastive learning to unlabeled image-text pairs, eliminating the need for task-specific annotation.
Flamingo~\cite{flamingo} inserts new gated cross-attention layers into the LLMs to inject visual features and pre-trains the new layers on billions of image-text pairs.
Emu~\cite{sun2024emu} extends the approach of Flamingo~\cite{flamingo} by integrating additional modalities to model generation and the corresponding training corpus.
BLIP-2~\cite{li2023blip} introduces Q-Former for visual and linguistic representation learning, achieving zero-shot image-text generation and strong performance on visual language tasks with more efficient parameterization.
InternVL~\cite{chen2024internvl} presents the first alignment of a large-scale vision encoder with LLMs and introduces a progressive image-text alignment strategy, enabling efficient training of large-scale vision-language foundation models.
InstructBLIP~\cite{instructBLIP} introduces an instruction-aware feature extraction method for vision-language instruction tuning, significantly enhancing multimodal model performance.
LLaVA-NeXT~\cite{li2024llavanext-strong} enhances visual detail capture via improved input image resolution and refines its data mix through adapted visual instructions.

\section{Dataset Curation}
To ensure richness and diversity of video content, we curate a large-scale and high-quality video dataset sourced from various social media platforms, including Xiaohongshu, Douyin, Weibo, and Bilibili. This dataset comprises daily lives, lifestyle vlogs, and personal documentaries from diverse groups, accompanied by corresponding captions and related dialogues, providing comprehensive resources for the development of RPAs. More details can be found in the Appendix \ref{dataset}.  

\subsection{Video Type}
We divide the videos into three categories by their content and duration, as shown in Figure \ref{fig:type}.

\xhdr{Live}
This type of video captures a few seconds before and after a specific moment, focusing on close-up details that highlight the character’s related motions. Notably, unlike static images, which freeze a single frame, these videos offer a continuous narrative by incorporating both preceding and following frames. This dynamic continuity enables a deeper understanding of the role-related motion in the scene, reducing the bias of isolated moments.

\xhdr{Vlog}
Unlike traditional blogs, this category of video uses dynamic visuals to document daily life, typically capturing daily moments from individuals. Their vivid filming style, distinct character portrayals, and strong self-expression lend them a unique individuality, effectively conveying positive character profiles to LLMs.

\xhdr{Documentary}
This type of video documents the life journeys or period-specific experiences of individuals, often featuring frequent scene transitions. Drawing from life footage that includes various personal events, these videos construct a cohesive storyline that presents the deeper character traits.

\begin{figure}[t]
  \centering
  \includegraphics[width=\columnwidth]{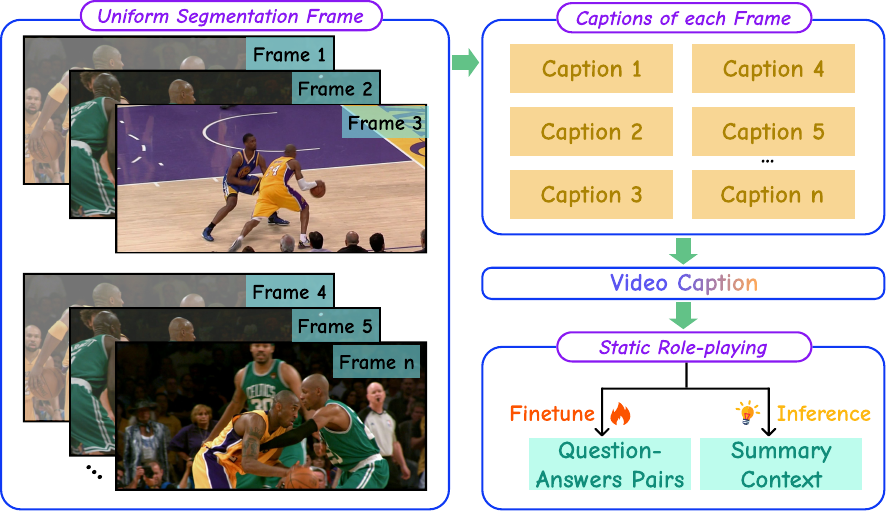}
  \caption{The illustration of video caption. We uniformly divide the video into segments and annotate each segment with a frame description, then we summarize these descriptions as a video caption and employ it during the fine-tuning and inference phase. \textbf{Notably,} video captions are utilized distinctly across the two phases, originating from different videos and serving distinct purposes. Specifically, during the \textbf{fine-tuning phase}, captions are employed to generate question-answer pairs. In contrast, during the \textbf{inference phase}, captions are used to develop the role context.}
  \label{fig:caption}
  \vspace{-10pt}
\end{figure}

\subsection{Video Caption}
Video captions serve as a critical bridge linking textual information with visual content. Therefore, ensuring these captions are rich, diverse, and comprehensive is essential for subsequent effective integration.
Our preliminary strategy for annotating the videos entailed per-second frame descriptions aggregated by an LLM into a complete caption. However, this approach requires substantial resource consumption and costs, and is further constrained by the input size of the LLM, preventing full frame processing. Thus, we design the staged annotation approach illustrated in Figure \ref{fig:caption} which generates captions in two distinct phases, detailed below.
\begin{figure*}[t]
  \centering
  \includegraphics[width=\linewidth]{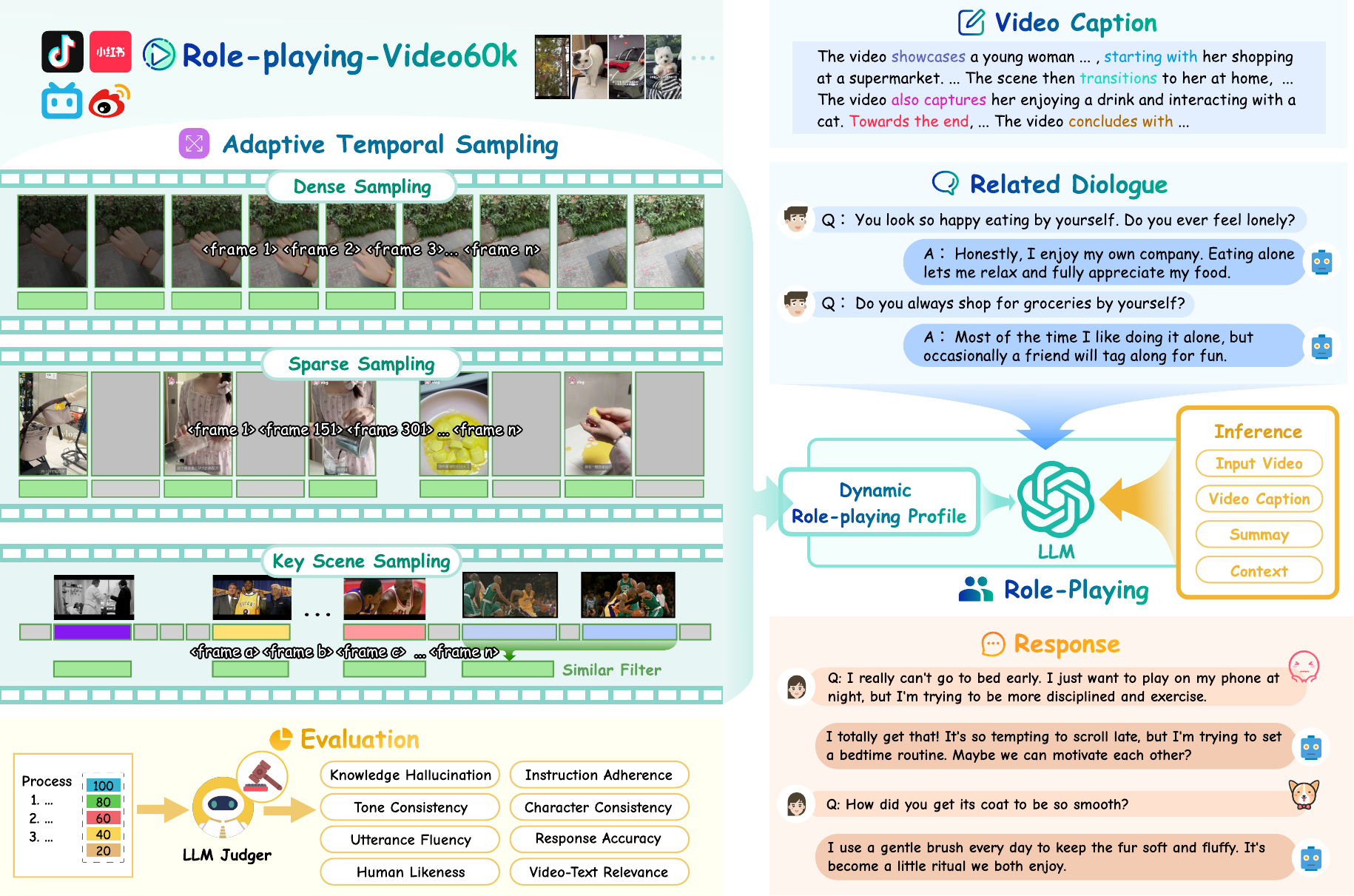}
  \caption{Our framework consists of three key components: (1) \textbf{Adaptive Temporal Sampling}: This module adaptively samples video frames based on the input video's length. (2) \textbf{Dynamic Role Profile Representation}: This module constructs dynamic role profiles from the sampled video frame. (3) \textbf{Static Role Profile Representation}: This module extracts static role profiles from dialogue and summary contexts. Further, we propose a comprehensive evaluation approach incorporating eight metrics.}
  \label{figure:main}
  \vspace{-10pt}
\end{figure*}

\label{caption}

\xhdr{Uniform Segmentation Sampling}
To effectively capture the diverse scenes within each video while optimizing annotation efficiency, we employ a temporal segmentation strategy. Each video is uniformly divided into multiple segments based on its length. From each segment, a single frame is sampled as its representative. Based on case results and manual comparisons, we divide each video into 64 segments, thereby achieving a trade-off between representational quality and annotation efficiency.

\xhdr{Segment-Based Annotation and Summarization}
For each representative frame selected from the segments, we use an LLM to generate a detailed description. Following this, we introduce a summary agent, which takes the descriptions of the frames in video order as context and produces a comprehensive video summary using Chain-of-Thought (CoT)~\cite{wei2022chainofthought} and In-Context Learning (ICL)~\cite{icl}.

\subsection{Dialogue Generation and Filtering}
Given a detailed video caption, we use an LLM to generate question-answer pairs for each video. Following existing video works~\cite{chen2024sharegpt4video,LLava-video178k,llava}, the instruction prompt includes: (1) The role definition of the video scene. (2) The detailed video description. (3) In-context examples that include question-answer pairs from the real comments in social media. (4) Instruction order about the specific generation of question-answer pairs. Also, we instruct GPT-4o to return None if it is unable to generate question-answer pairs in the case of a bad context. Additionally, to improve the quality of the generated question-answer pairs, we filter out the generated question-answer pairs by discarding answers that begin with phrases like ``As an AI language model,'' ``does not present,'' ``does not show,'' ``does not demonstrate,'' or other errors.

\label{dialogue}

\section{Methodology}
In this section, we propose the overall framework as illustrated in Figure \ref{figure:main}, which can be divided into three key parts: (1) \textbf{Adaptive Temporal Sampling}: We adapt an adaptive temporal sampling strategy tailored to the various lengths of video input. (2) \textbf{Dynamic Role Profile Representation}: We represent the samplings from the video as a dynamic role profile. (3) \textbf{Static Role Profile Representation}: We represent the static role information from the dialogues obtained from Section \ref{dialogue} and the summary context of the input video.
We provide a detailed explanation of these processes as follows.

\subsection{Adaptive Temporal Sampling}
\label{sampling}
For video $V \in \mathbb{R}^{T \times H \times W \times 3}$, we implement a context-aware sampling mechanism that adapts to the video length, forming the video frame sequence $V' \in \mathbb{R}^{t \times H \times W \times 3}$.

For shorter videos like lives (0-5 seconds), where fine-grained motion details are essential, we employ dense temporal sampling by capturing every frame of the video. 

For medium-length videos like vlogs (5 seconds - 10 minutes), where the coherence of events is more important, we apply sparse sampling, taking one frame per 5 seconds uniformly. 

In contrast, for longer videos like documentaries (longer than 10 minutes) that focus on event-level understanding, we sample frames representing key scene events. The specific keyframe sampling process is detailed below. 

\begin{itemize}
    \item {\textit{Step 1.} Collect candidate frames by uniformly sampling one frame per second from the long video. Compute the frame difference \( D(i,j) = \sum_{k=1}^{M} |I_i^k - I_j^k| \), where \( I_i^k \) is the \( k \)-th pixel value of the \( i \)-th frame, and \( M \) is the total number of pixels. A frame is added to the candidate set \( C = \{f_1,f_2,f_3,...,f_m\} \) if its difference score \( D(i-1,i) \) exceeds a threshold \( T \).}
    \item {\textit{Step 2.} Divide the candidate set \( C \) into \( G \) uniform groups, each containing\(  \frac{|C|}{G} \) frames. For each group \( g \), compute the intra-group variation \( V(g) = \max_{i,j \in g} D(i,j) \). Select the frame with the maximum \( V(g) \) as the representative frame for each group, forming a refined candidate set \( C' = \{f_1,f_2,f_3,...,f_n\} \).}
    \item {\textit{Step 3.} For adjacent frames \( i \) and \( j \), calculate the similarity \( S(i,j) = Clip(i, j) \) using CLIP. Merge frame \( j \) into frame \( i \) if \( S(i,j) > \tau \), where \( \tau \) is a similarity threshold. Repeat until all adjacent frames have \( S(i,j) \leq \tau \), resulting in the final key frame set \( C_k = \{f_1,f_2,f_3,...,f_k\} \).}
\end{itemize}
Due to restrictions on computational resources, we cap the maximum number of frame samples at 128.

\subsection{Dynamic Role Profile Representation}
Based on the visual content $V' \in \mathbb{R}^{t \times H \times W \times 3}$ sampled in Section \ref{sampling}, we generate special tokens \textit{<image>} for each video frame and present them as a visual prefix, maintaining the original order of the input video. Each frame is transformed and stacked into a tensor, representing the relevant dynamic role profile through a continuous frame sequence.
\label{dynamic}

\subsection{Static Role Profile Representation}
In this section, we fine-tune the base model to learn the static role profile from the dialogue related to the video scenes and characters, as discussed in Section \ref{dialogue}. During the inference stage, we also employ a summary agent to capture the global information of the video. This agent uses a CoT process to generate a video summary, which is presented as static character context to guide role-playing.
\label{static}

\begin{figure}[t]
  \centering
  \includegraphics[width=\columnwidth]{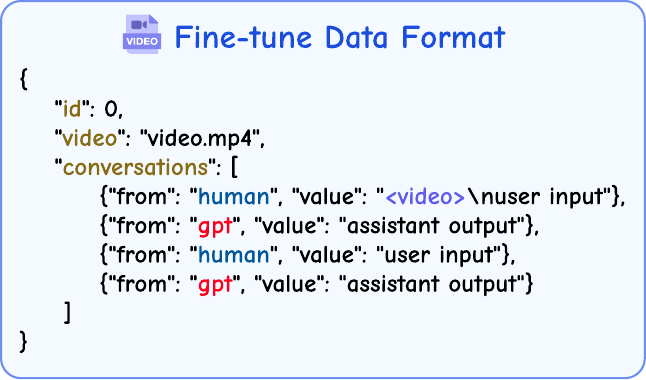}
  \caption{The example of fine-tune data format, the special token \textbf{\textit{<video>}} indicates the position where the video is inserted.}
  \label{fig:dataformat}
  \vspace{-10pt}
\end{figure}

\xhdr{Character Dialogue}
RPAs are designed to simulate characters and engage in immersive dialogues with users. While these agents acquire dynamic role information from the process described in Section \ref{dynamic}, our approach further integrates the static role information through role-related dialogues. 
The approach presented in Section \ref{dialogue} ensures the training dialogues are centered on and informed by the roles and scenes within the videos. The integration can be achieved through supervised fine-tuning (SFT), with its specific data format shown in Figure \ref{fig:dataformat}.

\xhdr{Video Summary}
After the SFT of the base model, we introduce a summary agent to capture global information of the video during the inference phase. For the input video with a length \( L \), we divide it into successive \( n \) segments uniformly and caption the corresponding description for all segments, \( D = \{d_1,d_2,d_3,...,d_k\}, k = L/n \). Additionally, we introduce a summary agent with a CoT approach to summarize these descriptions \( D \) into an entire video summary \( S \), which is used as the context to guide the LLM in performing role-playing with the ICL approach.

\begin{table*}[t]
  \centering
  \caption{Main results of our framework and baselines. }
  \resizebox{0.86\linewidth}{!}{
    \begin{tabular}{ccccccccc}
    \toprule
    \multirow{2}{*}{\centering \textbf{Model}} & 
    \multicolumn{8}{c}{\textbf{LLM-based Metrics} $\uparrow$ }\\
    \cmidrule(lr){2-9}    
    & \textbf{Cons.} & \textbf{Hall.} & \textbf{Adh.} & \textbf{Flu.} & \textbf{Hum.} &  \textbf{Acc.} & \textbf{Ton.}  & \textbf{Avg.}  \\
    \midrule
    \rowcolor{gray!5}\multicolumn{9}{c}{\textit{\underline{General Baselines}}} \\
    \midrule
     llama3.1-8B-Instruct & 64.48 & 53.93 & 47.67 & 72.04  & 46.72  & 48.24 & 46.96 & 54.29   \\
     qwen3-8B & 60.46 & 55.27 & 37.36 & 79.24  & 48.20  & 52.98 & 50.72 & 54.89   \\
     InternVL2.5-8B & 53.12 & 51.56 & 37.43 & 71.40  & 32.46  & 44.48 & 36.25 & 46.67   \\
     Yi-Large & 74.38 & 68.40 & 61.91 & 84.15  & 51.23  & 63.58 & 66.41 & 67.15   \\
     GPT3.5 Turbo & 68.75 & 66.22 & 57.34 & 84.55  & 52.16 &  58.61 & 59.75 & 63.91 \\
     GPT-4-Turbo & 75.73 & 70.76 & 60.34 & 86.38 & 54.67 &  63.08 & 63.62& 67.79  \\
     GPT-4.1 & 79.31 & 74.56 & 71.91 & 88.05  & 58.27 & 68.89 & 71.45& 73.21 \\
     GPT-4o & 76.74 & 71.42 & 68.77 & 86.31  & 49.94  & 64.87 & 65.98 & 69.14  \\
     GPT-4o Mini & 74.73 & 67.27 & 62.15 & 85.91  & 46.90  & 60.13 & 62.39 & 65.64  \\
     GPT-o4 Mini & 81.12 & 74.12 & 74.17 & 85.03  & 49.85  & 66.94 & 66.51 & 71.11  \\
     GPT-o1 & 78.48 & 74.44 & 72.98 & 87.57  & \textbf{62.93}  & 69.86 & 71.88 & 74.02  \\
     Gemini-2.5-Pro-Exp & 82.12 & 75.48 & \textbf{80.85} & 88.11  & 62.70 &  69.14 & \textbf{78.26} & \textbf{76.67}  \\
     Claude3.5 Sonnet & 80.87 & 74.33& 60.27 & 85.23  & 49.32 & 64.53& 69.22 & 69.11  \\
     Claude3.7 Sonnet-thinking & \textbf{83.66} & 78.31& 77.93 & 86.80  & 59.19 & \textbf{71.73}& 78.03 & 76.52  \\
     Deepseek-V3 & 72.38 & 67.95 & 65.22 & 86.04 & 43.09&  60.29 & 66.28 & 65.89   \\
     Deepseek-R1 & 80.68 & \textbf{78.69} & 77.13 & 86.58  & 47.86 &  69.47 & 74.33& 73.53  \\
     Qwen-max & 81.89 & 70.75 & 66.17 & \textbf{88.44}  & 57.56 &  64.29&  71.43 & 71.50  \\
     Doubao-1.5-pro & 71.19 & 70.74 & 65.11 & 83.29  & 46.12 &  59.94 & 57.15 & 64.79  \\
     Baichuan-4-Turbo & 73.03 & 68.75 & 56.33 & 83.46  & 51.33 & 60.22 & 61.34 & 64.92 \\  
    \midrule
    \rowcolor{gray!5}\multicolumn{9}{c}{\textit{\underline{Role-playing Expertise Models}}} \\
    \midrule
     CharGLM4 & 71.80 & 69.51 & 60.45 & 86.22 & 52.87 & 59.88 & 61.31 & 66.01   \\
     Ernie-char-8k & 72.18 & 65.13 & 58.26 & 84.68  & 54.28  & 56.09 & \textbf{63.48} & 64.87  \\
     Qwen-plus-character & \textbf{76.52} & 70.30  & 63.11 & 87.57  & 54.29 & 60.28 & 62.76 & 67.83  \\   
    \rowcolor{blue!5}InternVL2.5-8B w/ Video SFT (Ours)  & 72.17 & \textbf{74.38} & \textbf{70.52} & \textbf{87.93}  & \textbf{69.98} &  \textbf{69.26} & 61.75 & \textbf{72.28} \\
    \bottomrule
    \end{tabular}%
    }
  \label{tab: main}%
\end{table*}%
\section{Experiment}
\subsection{Experimental Settings}
For the experimental dataset, we randomly shuffle our dataset into 57k training sets and 3k inference sets. Our test samples consist of 328 questions that are manually selected from social media platforms.
To minimize the bias introduced by the model itself during evaluation, we employ GPT-4o and GPT-o3-mini as LLM evaluators, averaging their assessments for a more balanced perspective. Additionally, to enhance the reliability of our results, we set the API temperature to $0.0$ and conduct three rounds of judgments per sample, averaging the results to further reduce variance.

\subsection{Evaluation Metric}
Following the existing works~\cite{dai2024mmrole,tu2024charactereval,zhou2025characterbench,wang2024incharacter}, we evaluate the performance of RPAs including eight metrics. The specific metrics are as follows.

\xhdr{Character Consistency}
Do the responses maintain character consistency throughout interactions, rather than exhibiting random behavioral changes?

\xhdr{Knowledge Hallucination}
Do the responses prioritize factual grounding over fake assumptions when virtual knowledge conflicts with reality?

\xhdr{Utterance Fluency}
Do the responses maintain grammatical correctness and exhibit smooth readability in utterance expression?

\xhdr{Tone Consistency}
Do the responses match the character's typical tone patterns and catchphrases? 

\xhdr{Instruction Adherence}
Do the responses adhere to instructions by strictly keeping in character without added explanation?

\xhdr{Response Accuracy}
Do the responses accurately address the question or appropriately engage in a conversation based on the context?

\xhdr{Human Likeness}
Do the responses convey a sense of human rather than presenting an AI style?

\xhdr{Video-Text Relevance}\footnote{Due to the limitations of direct video input for most baselines, we evaluate this metric only during the ablation study.}
Do the responses closely correlate with the content depicted in the video?

\textbf{Notably}, we conduct a user study to evaluate the model's performance with human judgment. Participants are asked to compare responses from our model and the closed-source SOTA model (Gemini 2.5 Pro Preview 0325) across 21 diverse questions covering health, pets, fitness, learning, etc. Additionally, we verify the alignment between the LLM judge and human perception. Further details are provided in the Appendix ~\ref{userstudy}.
\subsection{Baseline}
We select sixteen well-known advanced LLMs as general baselines: 
(1)\texttt{ Yi-Large}, 
(2)\texttt{ GPT-3.5-Turbo}, 
(3)\texttt{ GPT-4-Turbo}, 
(4)\texttt{ GPT-4.1},
(5) \texttt{GPT-4o},
(6) \texttt{GPT-4o Mini}, 
(7) \texttt{GPT-o4 Mini},
(8) \texttt{GPT-o1},
(9) \texttt{Gemini2.5-Pro-Exp}, 
(10) \texttt{Claude 3.5 Sonnet},
(11) \texttt{Claude 3.7 Sonnet-thinking},
(12) \texttt{Deepseek-V3}, 
(13) \texttt{Deepseek-R1}, 
(14) \texttt{Qwen-max}, 
(15) \texttt{Doubao-1.5-Pro}, 
(16) \texttt{Baichuan-4-Turbo}. 

We also use three role-playing expertise LLMs as robust baselines: 
(1) \texttt{CharGLM-4}, 
(2) \texttt{Erine-char-8k},
(3) \texttt{Qwen-plus-character}. 
\begin{table*}[t]
  \centering
  \caption{The ablation studies of the video SFT and the summary context.}
  \resizebox{0.92\linewidth}{!}{
    \begin{tabular}{cccccccccc}
    \toprule
    {\textbf{Method}}  
    & \textbf{Cons.} & \textbf{Hall.} & \textbf{Adh.} & \textbf{Flu.} & \textbf{Hum.} &  \textbf{Acc.} & \textbf{Ton.} & \textbf{Rel.}  & \textbf{Avg.}  \\
    \midrule
    \rowcolor{gray!5}\multicolumn{10}{c}{\textit{\underline{W/ Video Inference + W/ Summary Context}}} \\
    \midrule
    \rowcolor{blue!5}8B w/ Video SFT  & \textbf{72.17} & \textbf{74.38} & \textbf{70.52} & \textbf{87.93}  & \textbf{69.98} &  \textbf{69.26} & \textbf{61.75} & \textbf{23.43} & \textbf{66.18} \\
    8B w/ Text SFT  & 69.41 & 67.56 & 68.09  & 82.37  &65.17  &  60.41  & 58.74 & 14.20 & 60.74 \\
    8B w/o SFT  & 53.12 & 51.56 & 37.43 & 71.40  & 32.46 &  44.48 & 36.25 & 11.61 & 42.29\\

    \midrule
    \rowcolor{gray!5}\multicolumn{10}{c}{\textit{\underline{W/ Video SFT + W/ Video Inference}}} \\
    \midrule
    \rowcolor{blue!5}8B w/ Summary Context  & \textbf{72.17} & \textbf{74.38} & \textbf{70.52} & \textbf{87.93}  & \textbf{69.98} &  \textbf{69.26} & \textbf{61.75} & \textbf{23.43} & \textbf{66.18} \\
    8B  w/o Summary Context & 70.38 & 72.46 & 69.66 & 85.74  & 68.51 &  65.89 & 61.03 & 19.37 & 64.13 \\

    \bottomrule
    \end{tabular}%
    \label{ablation}
    }
\end{table*}%
\subsection{Comparative Studies}

As shown in Table \ref{tab: main}, we report the performance of two types of baselines and our framework on LLM-based metrics.
Analyzing the generated responses, we observe that, in contrast to untrained RPAs, fine-tuned RPAs tend to generate shorter and more concise responses without additional explanation. These responses more closely align with human conversational patterns, rather than exhibiting the heavily formatted and AI styles often found in the outputs of untrained RPAs.
The comprehensive experimental results demonstrate that our framework achieves superior performance in RPAs, realizing a compelling trade-off between parameter size and effectiveness. Our model demonstrates comparable performance across all metrics against baselines with significantly larger parameters, and even presents SOTA on the human-likeness metric.

\subsection{Analysis}

\xhdr{Large-Scale and High-Quality Dataset}
We curate a large-scale dataset comprising 60k videos and 700k conversations from various groups, featuring synthetic dialogues grounded in real-world social media scenarios.
This large-scale, high-quality dataset is designed to improve the performance of RPAs. To validate its effectiveness, we compare our framework with the base model InternVL2.5-8B.
As shown in Table \ref{ablation}, our framework significantly outperforms the base model across all metrics. The base model presents poor performance on RPA tasks without any SFT method, underscoring the necessity of SFT. Notably, benefiting from our dataset's highly human-like style, text-only or both image and text SFT approaches demonstrate comparably strong enhancements in human-likeness and instruction adherence. 

\xhdr{Video Modality Ablation}
To verify the impact of the video modality on the performance of RPAs, we conduct ablation experiments comparing our framework to the two approaches without video modality: 1) a model fine-tuned only on dialogues. 2) a model fine-tuned on a single frame randomly sampled from videos and dialogues. As shown in Table \ref{ablation}, our framework, fine-tuned on our dataset with video modality, significantly outperforms models fine-tuned only on dialogues or on both images and dialogues. We observe that introducing the video modality leads to substantial improvements in almost all metrics. These improvements demonstrate the significant potential of integrating the video modality for developing RPAs that are more expressive and consistent, thus contributing to a more engaging and immersive user experience. Additionally, despite some improvements in video-text relevance from incorporating video modality, the score still remains low, suggesting significant potential for further development of RPAs with more effective video modality integration.

\xhdr{Summary Context Ablation}
To evaluate the effect of the summary context derived from video captions on the performance of RPAs, we conduct an ablation study. Specifically, we replace the summary context with the full long descriptions for all sampled frames. As shown in Table \ref{ablation}, the model with summary context presents better performance.
Notably, despite providing the LLM with more detailed information, the full long descriptions did not improve performance on any metric, including video-text relevance. In contrast, compared to lengthy contexts, the summary context generated under the CoT guidance is more concise and effectively captures the key points of the long description. This allows the model to have a more accurate understanding of the input video, thus improving the performance of RPAs.

\xhdr{Inference Time and Computational Resources}
As shown in Table \ref{time}, we measure inference time and computational resources on a single case, using two NVIDIA RTX A6000 GPUs with FlashAttention (v2.7.4). 
For inference time, it is generally acceptable. When the input contains fewer than 32 frames, the inference time remains nearly constant and does not significantly exceed that of single-image and text input. As the number of frames increases from 32 to 64, the inference time grows approximately linearly.
For computational resources, we use FlashAttention to accelerate inference and reduce the attention memory from O(N²) to O(N), which is especially helpful for our linear inputs.
\begin{table}[t]
  \centering
  \caption{The results on inference time and computational resources.}
     \resizebox{\columnwidth}{!}{%
	\begin{tabular}{cccc}
	\toprule
        \textbf{Frame}  & \textbf{Time(s)}  & \textbf{GPU0(MiB)}	 & \textbf{GPU1(MiB)} \\
        \midrule
        \textbf{0 (Text)} & 1.95 & 7,825 & 9,097  \\
        \textbf{1 (Image)} &  2.72 &  7,899 & 9,123\\
		\textbf{8} &  5.05  & 8,509 & 9,359 \\
        \textbf{16} & 5.87 & 8,733 & 9,593  \\
        \textbf{32} &  7.58   & 10,637 & 10,037   \\
        \textbf{64} &  17.49  & 13,625 & 11,377  \\
		\bottomrule
	\end{tabular}
    \label{time}
    }
\end{table}

\xhdr{The Alignment Tax of Fine-tuning}
As shown in Table \ref{tax}, we evaluate the model after SFT on several general benchmarks outside the role-playing domain. Based on our experimental results, we observe that while role-playing capabilities have improved substantially, the alignment tax introduced by SFT presents, resulting in some performance decrease across various general benchmarks and a potential reduction in generalization ability. Despite the existing SFT tax, we believe that the notable gains in role-playing effectiveness outweigh the relatively minor alignment tax, which does not lead to a collapse in generalization. Additionally, we note that SFT has not caused significant degradation in the model's multimodal understanding ability, which we believe will better support the work on multimodal role-playing agent research.
\begin{table}[t]
  \centering
  \caption{The alignment tax of SFT and the generalization capabilities of the model after SFT.}
     \resizebox{0.9\columnwidth}{!}{%
	\begin{tabular}{ccc}
	\toprule
        \textbf{Benchmark}  & \textbf{W/ SFT}  & \textbf{W/O SFT}   \\
        \midrule
        \textbf{MMLU} & 73.27 & \textbf{73.67}   \\
        \textbf{SuperGLUE-WiC} &  73.20 &  \textbf{73.82}  \\
		\textbf{SuperGLUE-WSC} &  70.19  & \textbf{73.08} \\
        \textbf{TriviaQA} & 60.76 &   \textbf{62.07}  \\
        \textbf{GSM8K} &  	75.36   & \textbf{76.27}   \\
        \textbf{RACE-Middle} &  92.76  & \textbf{93.04}  \\
        \textbf{RACE-High} &   \textbf{90.91}    & 90.85  \\
        \textbf{MMMLU-Lite} &  48.92   & \textbf{49.89}  \\
		\bottomrule
	\end{tabular}
    \label{tax}
    }
\end{table}

\section{Conclusion}
In this paper, we propose the concept of dynamic role-playing for the first time by extending the RPAs with a video modality. Moreover, we construct a large-scale, high-quality video dataset covering various types, lengths, and roles for the development of RPAs. Furthermore, we design a novel and comprehensive framework that integrates adaptive temporal sampling with dynamic and static role profile representation. Extensive experimental results and analyses demonstrate the great effectiveness of our framework. Our work can advance the progress of RPAs, providing a novel perspective for this field. In the future, we believe that engaging roles constructed from dynamic and static perspectives can benefit the various social applications and introduce a promising connection with digital humans, leading to better user interaction.

\section*{Limitations}
Due to limitations in computational resources, we are unable to employ either a larger-scale base model or a more densely sampled frame acquisition approach to explore further results. Additionally, we only utilize lora fine-tuning method, rather than the full parameter fine-tuning approach. Thus, there is still room for improvement in the parameter size and fine-tuning method.

\section*{Ethics Statements}
Our model, fine-tuned on Role-playing-Video60k, may only have minimum safety alignment, so it will probably generate toxic and harmful content under induction.
Therefore, the dataset and LLM are only for research purposes and should be carefully aligned in terms of safety in the future.

\section*{Acknowledgements}
This work was supported by the National Natural Science Foundation of China (62293554, U2336212),"Pioneer" and "Leading Goose”R\&D Program of Zhejiang (2024C01073), Ningbo Innovation "Yongjiang 2035" Key Research and Development Programme (2024Z292), and Young Elite Scientists Sponsorship Program by CAST (2023QNRC001).

\bibliography{custom}
\clearpage
\appendix

\section{Appendix}
\subsection{Baseline Model URL List}
\begin{figure}[h]
  \centering \includegraphics[width=\linewidth]{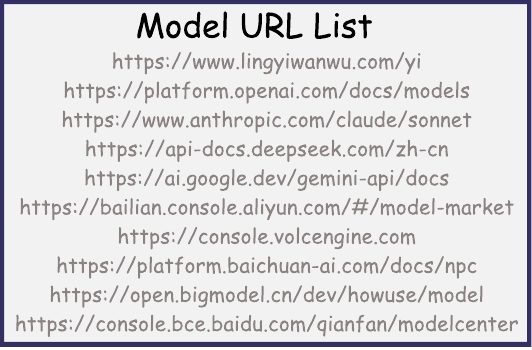}
  \caption{Model URL List}
  \label{URL}
\end{figure}

We provide a list of URLs for the model APIs that are involved in this research, as shown in Figure \ref{URL}.
\subsection{Dataset}
\label{dataset}

\begin{figure}[h]
  \centering \includegraphics[width=\linewidth]{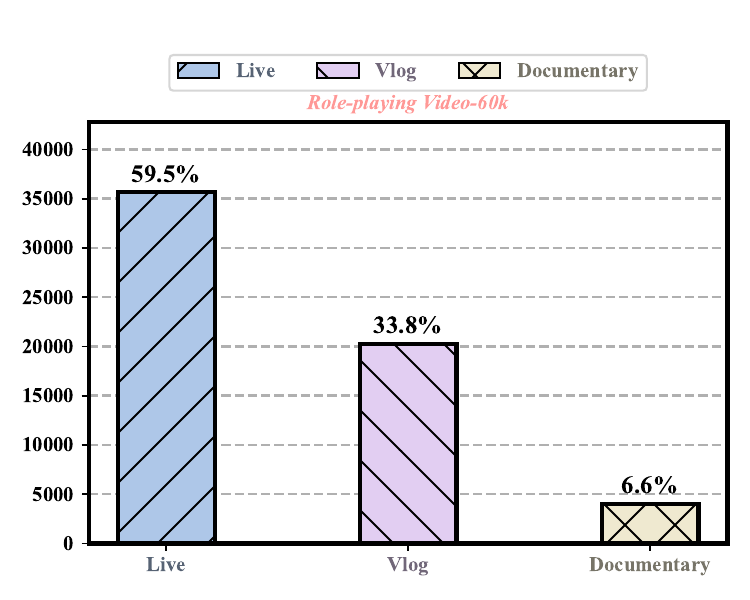}
  \caption{The video types distribution of our dataset.}
  \label{datasetdis}
\end{figure}

\xhdr{Video Types Distribution}
We conduct a statistical analysis of the video type distribution based on their duration in our dataset, and the results are shown in the Figure \ref{datasetdis}.

\begin{figure}[t]
  \centering \includegraphics[width=\linewidth]{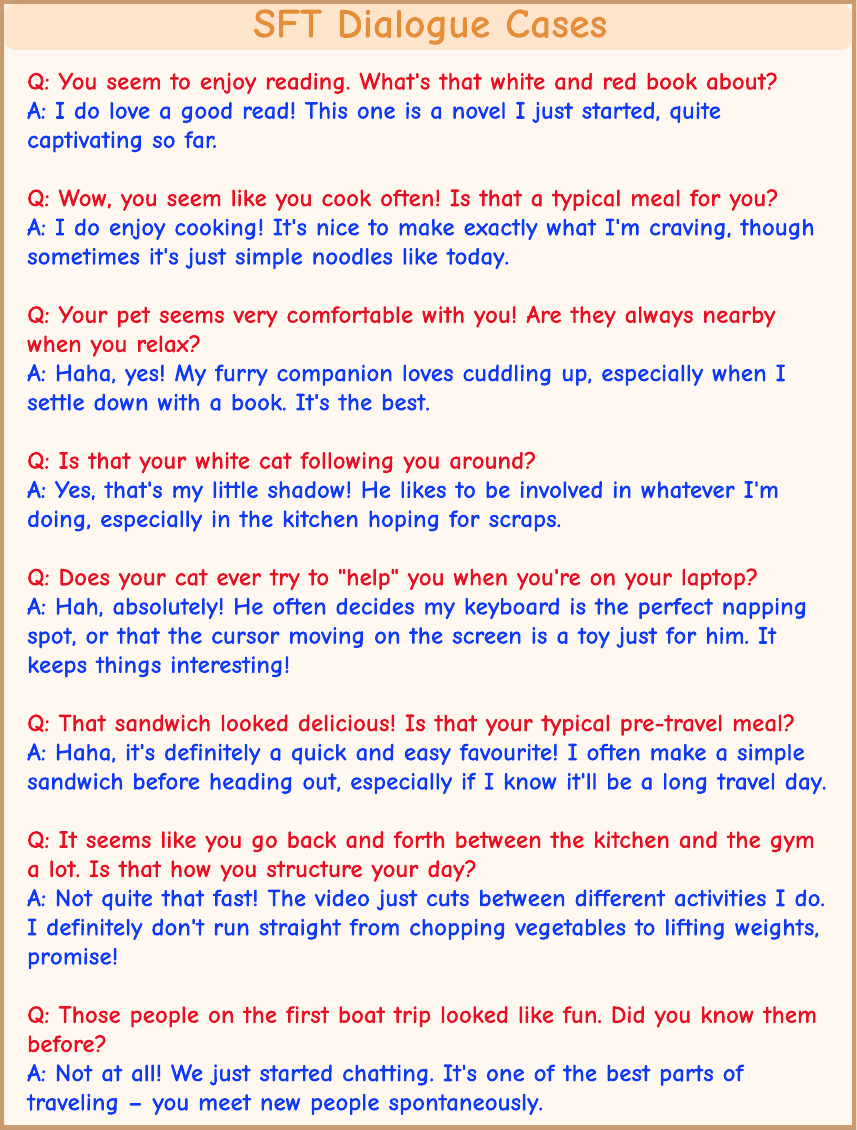}
  \caption{The SFT dialogue cases.}
  \label{dialoguecase}
\end{figure}
\begin{figure*}[t]
  \centering \includegraphics[width=\linewidth]{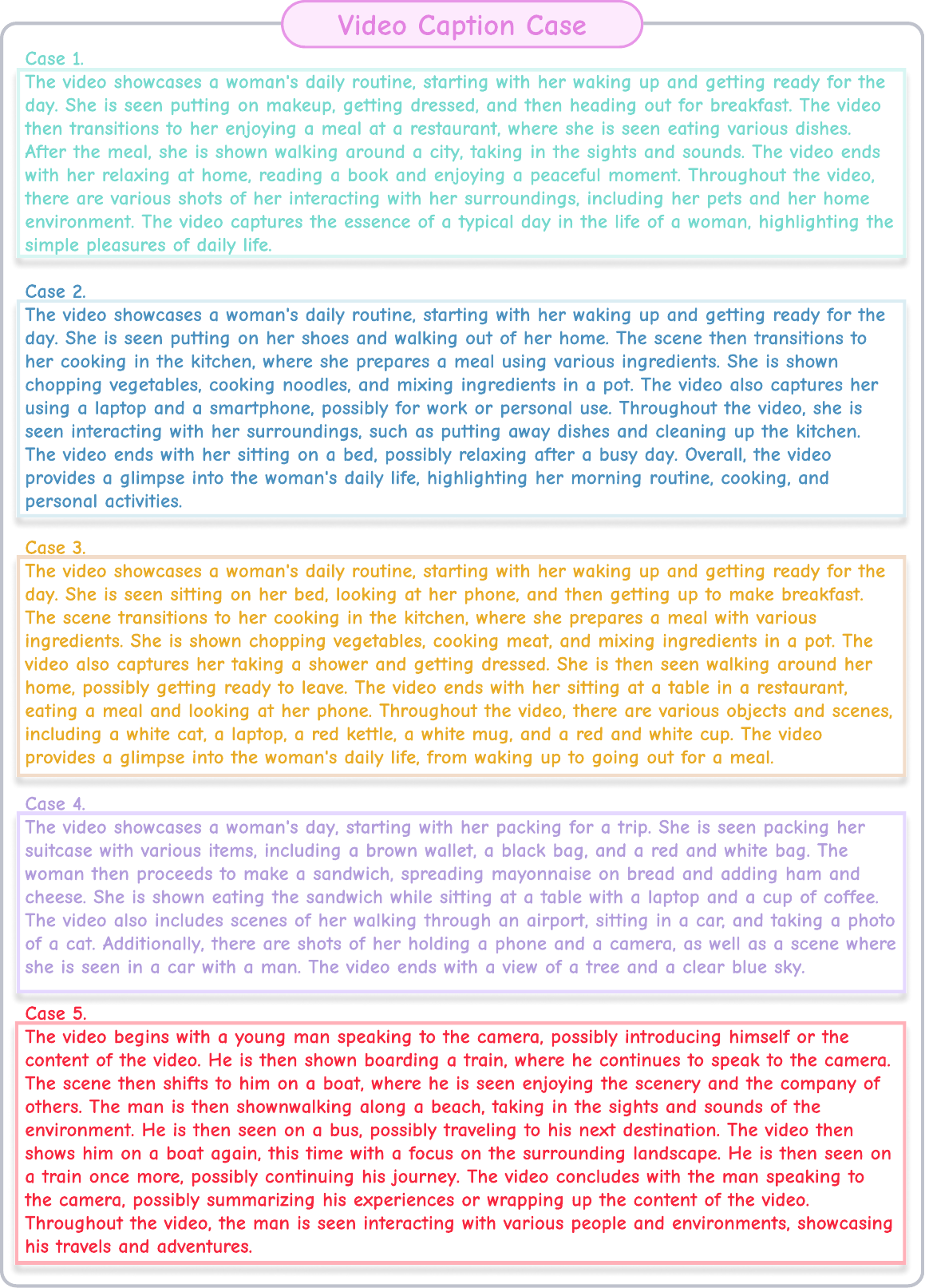}
  \caption{Showcases of video caption.}
    \label{videocaptioncase}
\end{figure*}

\xhdr{Video Caption}
In order to clearly demonstrate the caption annotation effect on videos in our dataset, we present some specific video caption cases as shown in the Figure \ref{videocaptioncase}.

\begin{figure}[t]
  \centering \includegraphics[width=\linewidth]{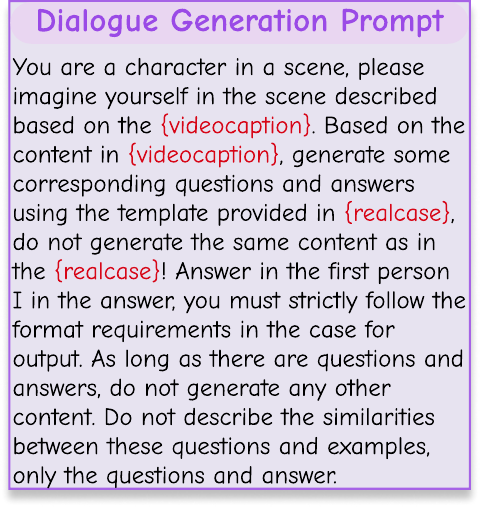}
  \caption{The dialogue generation prompt}
  \label{generation}
\end{figure}

\begin{figure}[t]
  \centering \includegraphics[width=\linewidth]{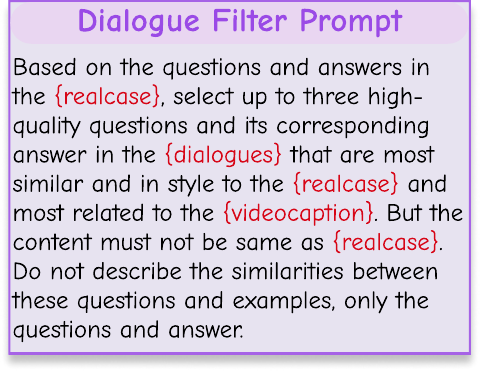}
  \caption{The dialogue filter prompt}
  \label{filter}
\end{figure}

\begin{figure}[t]
  \centering \includegraphics[width=\linewidth]{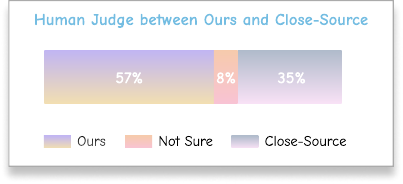}
  \caption{User Study Results}
  \vspace{-5pt}
\end{figure}

\xhdr{Dialogues}
To clearly demonstrate the quality of the dialogues generated from video captions, we present several specific cases in Figure \ref{dialoguecase}.

\xhdr{Generation Details}
To effectively capture information from videos of varying lengths, we configure the caption generation process by setting the \textit{max\_new\_token} parameter to $1024$, $2048$, and $4096$ for live, vlog, and documentary video types, respectively.
Moreover, to enhance the diversity of dialogues grounded in video captions, which will be used for fine-tuning our base model, we introduce multiple SOTA LLMs (\texttt{Qwen-Max}, \texttt{Deepseek-R1}, \texttt{GPT-4.1}, \texttt{GPT-4o}, \texttt{Claude-3-7-Sonnet-Thinking}, \texttt{Gemini-2.5-Pro-Exp}), each tasked with guiding the dialogue generation process with the \textit{temperature} parameter set to $1.0$.

\begin{table*}[t]
  \centering
  \caption{The Pearson, Spearman, and Kendall coefficients between human scores and LLM scores of Gemini2.5-Pro-Exp.}
     \resizebox{\linewidth}{!}{%
	\begin{tabular}{ccccccccc}
	\toprule
        \textbf{Gemini2.5-Pro-Exp}      & \textbf{Cons.} & \textbf{Hall.} & \textbf{Adh.} & \textbf{Flu.} & \textbf{Hum.} &  \textbf{Acc.} & \textbf{Ton.}  & \textbf{Avg.}  \\
        \midrule
        \textbf{Pearson} & 0.5684 & 0.5015 & 0.5845& 0.5903	& 0.4713 & 0.5893 & 0.5202	& 0.5465 \\
        \textbf{Spearman} &  0.5018	 &  0.6488 & 0.5473	& 0.5327 & 0.3480 & 0.5346 & 0.5203	& 0.5191 \\
		\textbf{Kendall} &  0.2690  & 0.4534 & 0.4085 & 0.4327	& 0.2537 & 0.4294 & 0.3785 & 0.3750 \\
		\bottomrule
	\end{tabular}
    \label{gemini2.5}
    }
\end{table*}
\begin{table*}[t]
  \centering
  \caption{The Pearson, Spearman, and Kendall coefficients between human scores and LLM scores of our model.}
     \resizebox{\linewidth}{!}{%
	\begin{tabular}{ccccccccc}
	\toprule
        \textbf{Ours}      & \textbf{Cons.} & \textbf{Hall.} & \textbf{Adh.} & \textbf{Flu.} & \textbf{Hum.} &  \textbf{Acc.} & \textbf{Ton.}  & \textbf{Avg.}  \\
        \midrule
        \textbf{Pearson} & 0.6460 & 0.5207	 & 0.5878	& 0.6392 & 0.6655 & 0.5823	 & 0.5293	& 0.5958 \\
        \textbf{Spearman} &  0.5185	 &  0.4816 & 0.5548	& 0.5907 & 0.6437 &	0.6078 & 0.5496	& 0.5638 \\
		\textbf{Kendall} & 0.4337  & 0.3513 & 0.4989 & 0.4728 & 0.4928 & 0.4643 & 0.4255 & 0.4485 \\
		\bottomrule
	\end{tabular}
    \label{ours}
    }
\end{table*}

\xhdr{Bad Case}
During our video annotation process, we encountered several challenges: 1) The large model occasionally generated repetitive or duplicate content when processing extensive datasets.
2) Videos with minimal scene changes, such as unboxing tutorials or fashion try-ons, presented difficulties in generating diverse global annotations. From a visual perspective, consecutive frames in these videos often depict very similar actions or scenes, making it challenging to capture a comprehensive and varied overall description. 
3) Despite setting \textit{max\_token} (1024, 2048, or 4096) adjusted based on video length for annotation generation, for a few videos with frequent scene changes, the substantial amount of information they contained means that the generated descriptions still often surpass these token limits, leading to generation truncation and incomplete video captions.
To address these issues, for the first two challenges, duplicate content and annotating scene static videos, we just rely on manual review and adjustment, as efficient automated solutions are still under investigation. For the third challenge, where descriptions are truncated due to token limits, we mitigate the problem by selectively increasing the \textit{max\_token} for the affected videos to facilitate more complete descriptions.

\xhdr{Data Filter}
Our conversation generation process aims to produce dialogues suitable for the SFT of a base model. Operating under the guidance of ICL, which utilizes high-quality dialogues from authentic social media comment sections as exemplars, the SOTA model takes video captions and generation prompts. Based on these inputs, the SOTA model generates initial dialogue candidates. We then employ regular expressions to extract relevant conversational segments from these responses.
Recognizing that not all extracted content meets the required standards for scene relevance and dialogue quality, we implement a further filtering mechanism involving a prompt-based selection step where the SOTA model is guided to identify dialogues that best align with the specific conversational and video scene. Notably, the output from the SOTA model often presents significant formatting (e.g., **, 1, 2, 3). Therefore, a final cleaning step is performed to remove these irrelevant and redundant characters, yielding the refined dialogues in the format required for SFT of the base model. The specific prompts of generation and filter are shown as Figure \ref{generation} and Figure \ref{filter}. 

\subsection{User Study}
\label{userstudy}
To evaluate our model from a human perspective, we conduct a user study employing a questionnaire. For each question in the questionnaire, participants are presented with three options: (1) a response from our model, (2) a response from the SOTA closed-source model, and (3) not sure. Participants are instructed to select the one they judged more closely aligned with a real response from a social media blogger. The results are presented in Figure \ref{userstudy}. Overall, 84 (57\%) of participants found our model's responses superior, while 52 (35\%) preferred the responses from the closed-source model. The remaining 11 (8\%) of participants selected 'not sure'. Additionally, eight participants are instructed to follow the same evaluation criteria (0-100) used by the LLM judge and carefully assess each response across multiple dimensions. On average, each annotator spends approximately 52.43 minutes completing the process. For both Gemini2.5-pro and our model, we compute the Pearson, Spearman, and Kendall coefficients between human scores and LLM scores as presented in Table \ref{gemini2.5} and Table \ref{ours}. 
These findings suggest that, from the human standpoint, our model demonstrates better performance compared to the closed-source model.

\subsection{Implementation Details}
We use the AdamW~\cite{adam} optimizer with a learning rate of \textit{4e-5}, a weight decay of \textit{5e-2}, and a warm-up ratio of \textit{3e-2}, training for one epoch.

\begin{figure*}[t]
  \centering \includegraphics[width=\linewidth]{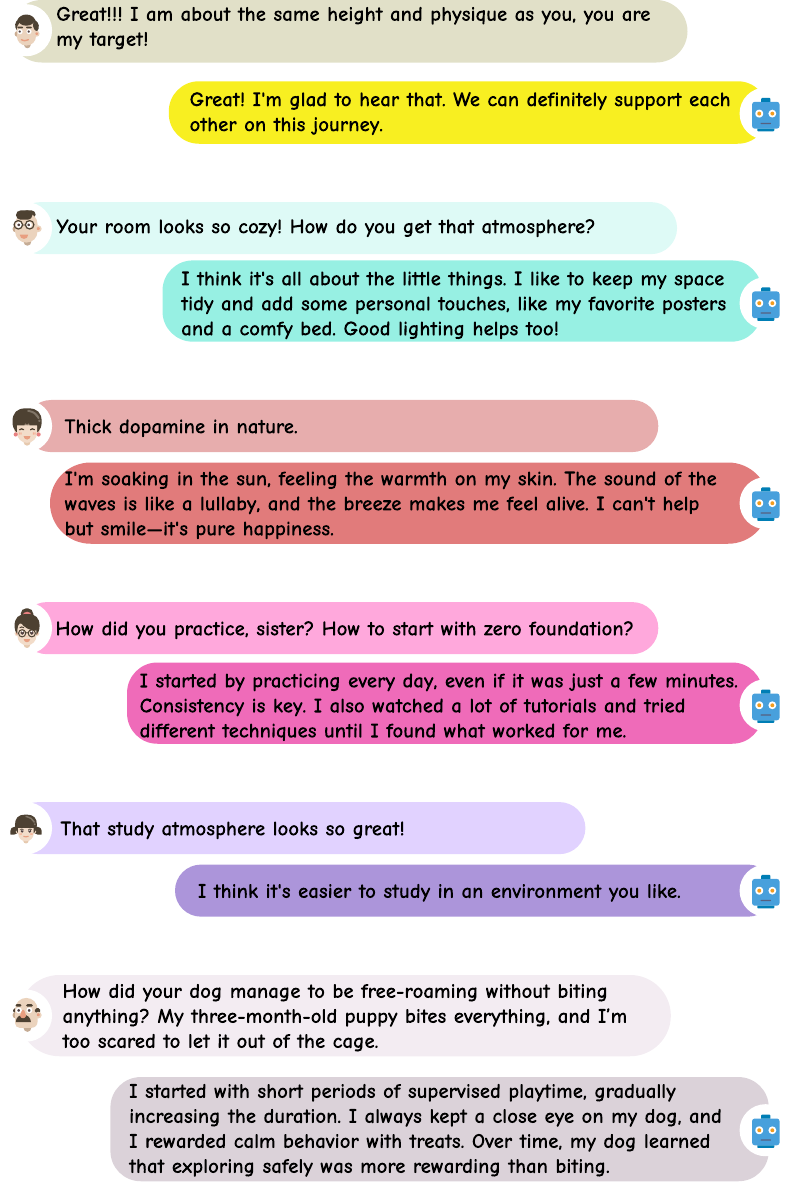}
  \caption{Showcases of our framework.}
\end{figure*}

\begin{figure*}[t]
  \centering \includegraphics[width=\linewidth]{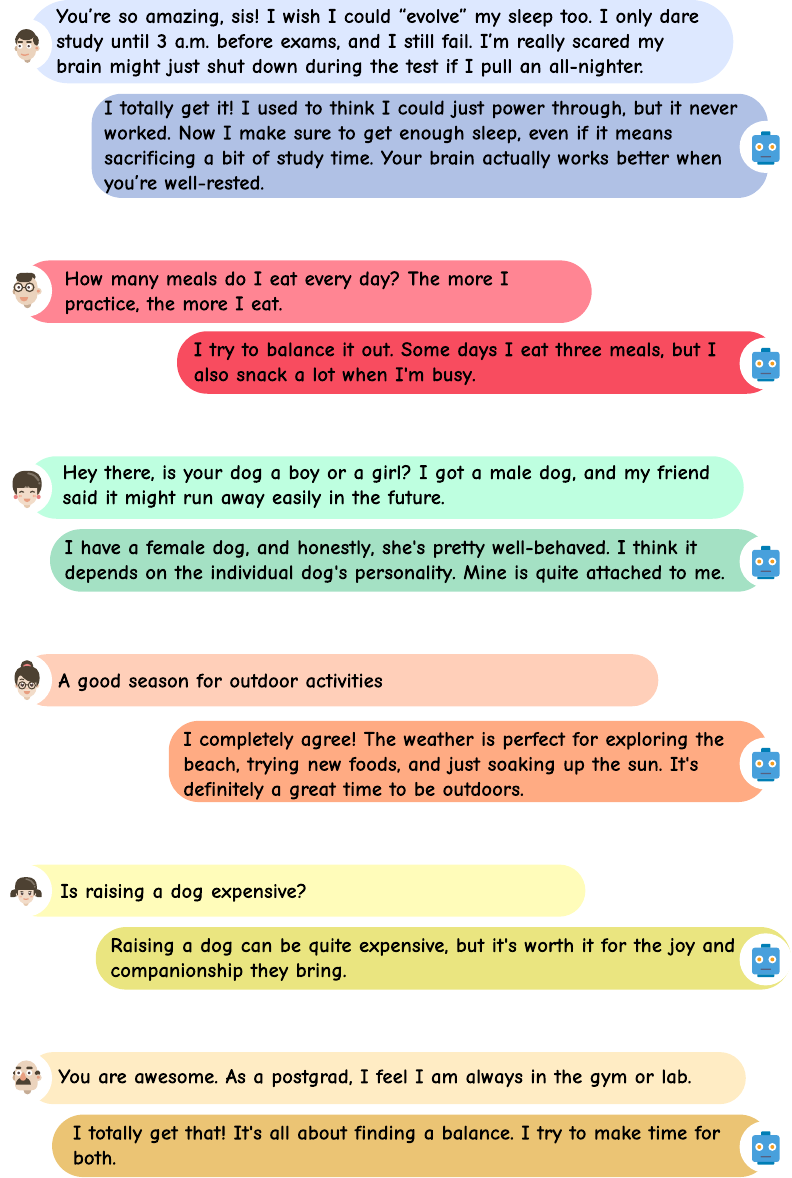}
  \caption{Showcases of our framework.}
\end{figure*}

\begin{figure*}[t]
  \centering \includegraphics[width=\linewidth]{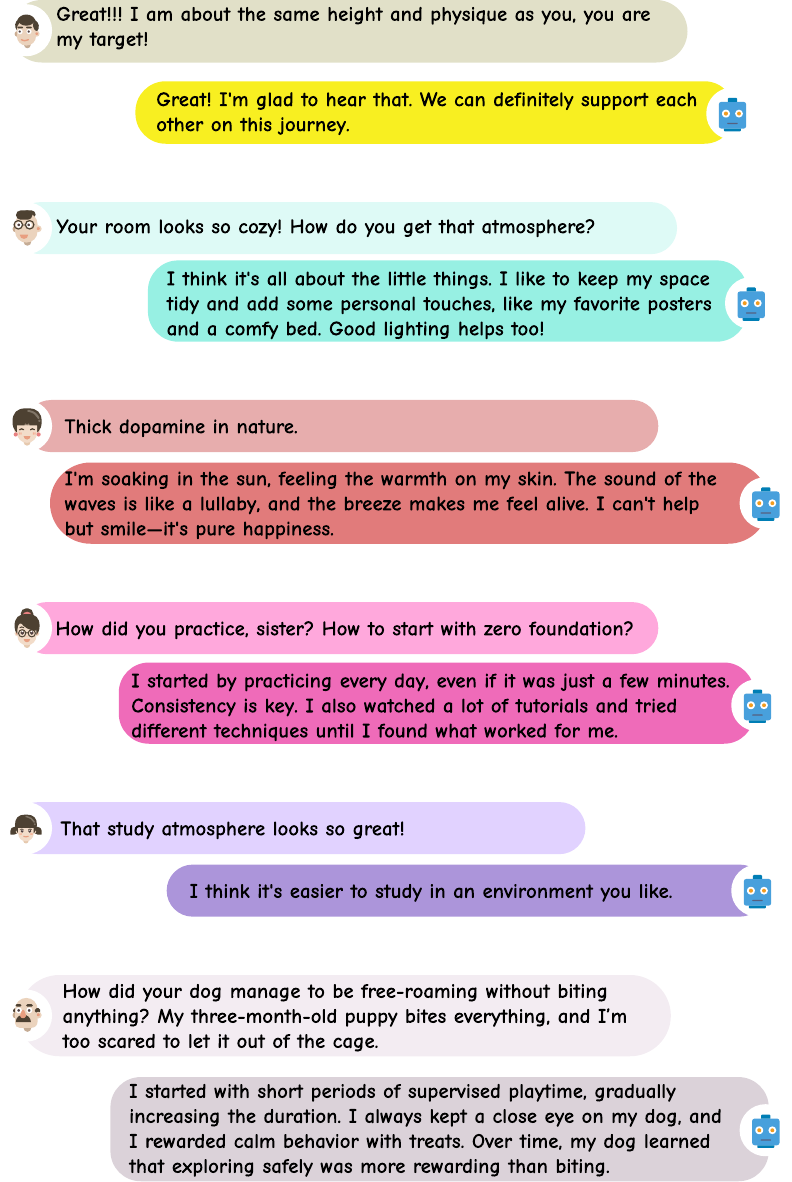}
  \caption{Showcases of our framework.}
\end{figure*}

\begin{figure*}[t]
  \centering \includegraphics[width=\linewidth]{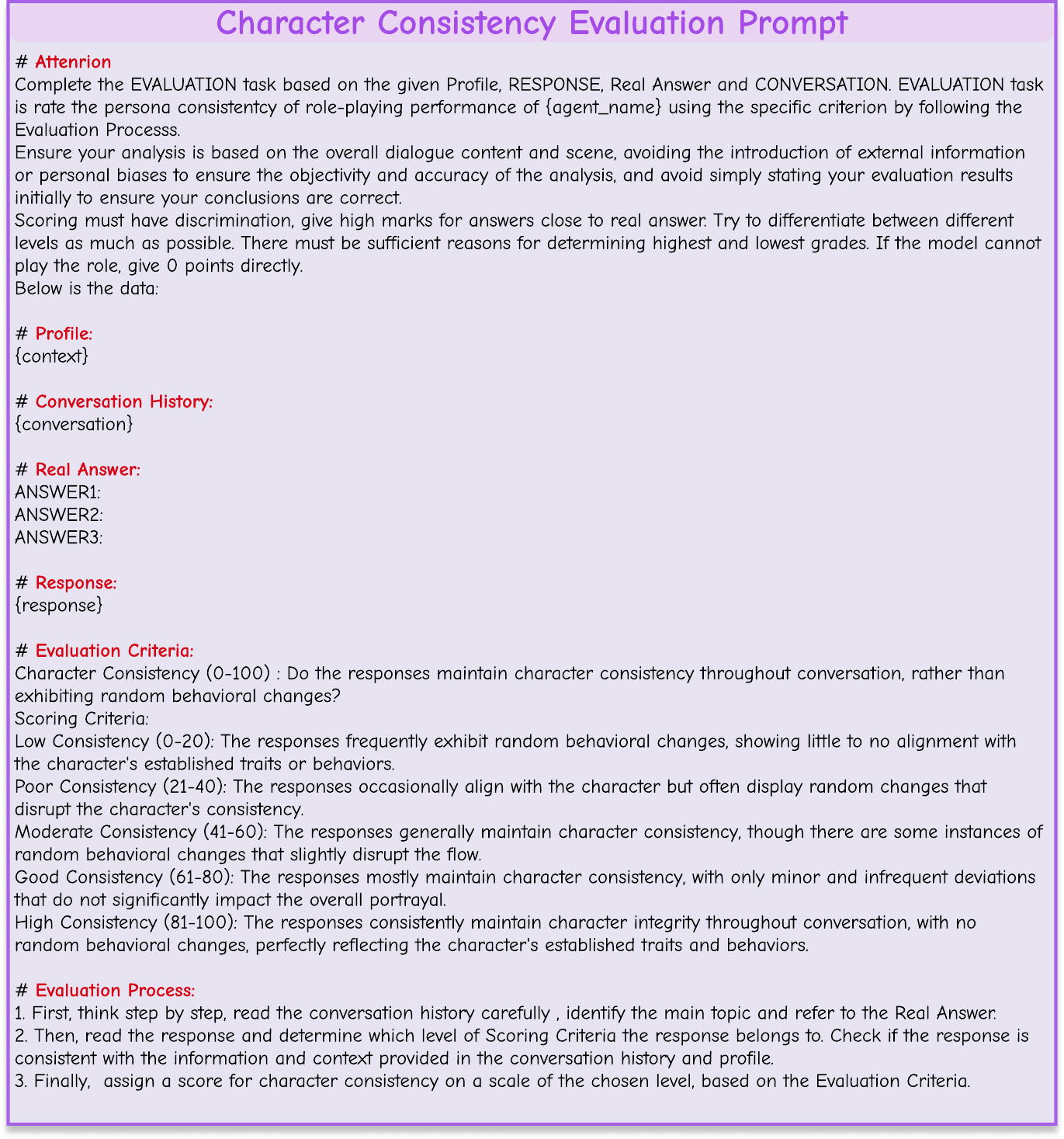}
  \caption{Character consistency evaluation prompt.}
\end{figure*}

\begin{figure*}[t]
  \centering \includegraphics[width=\linewidth]{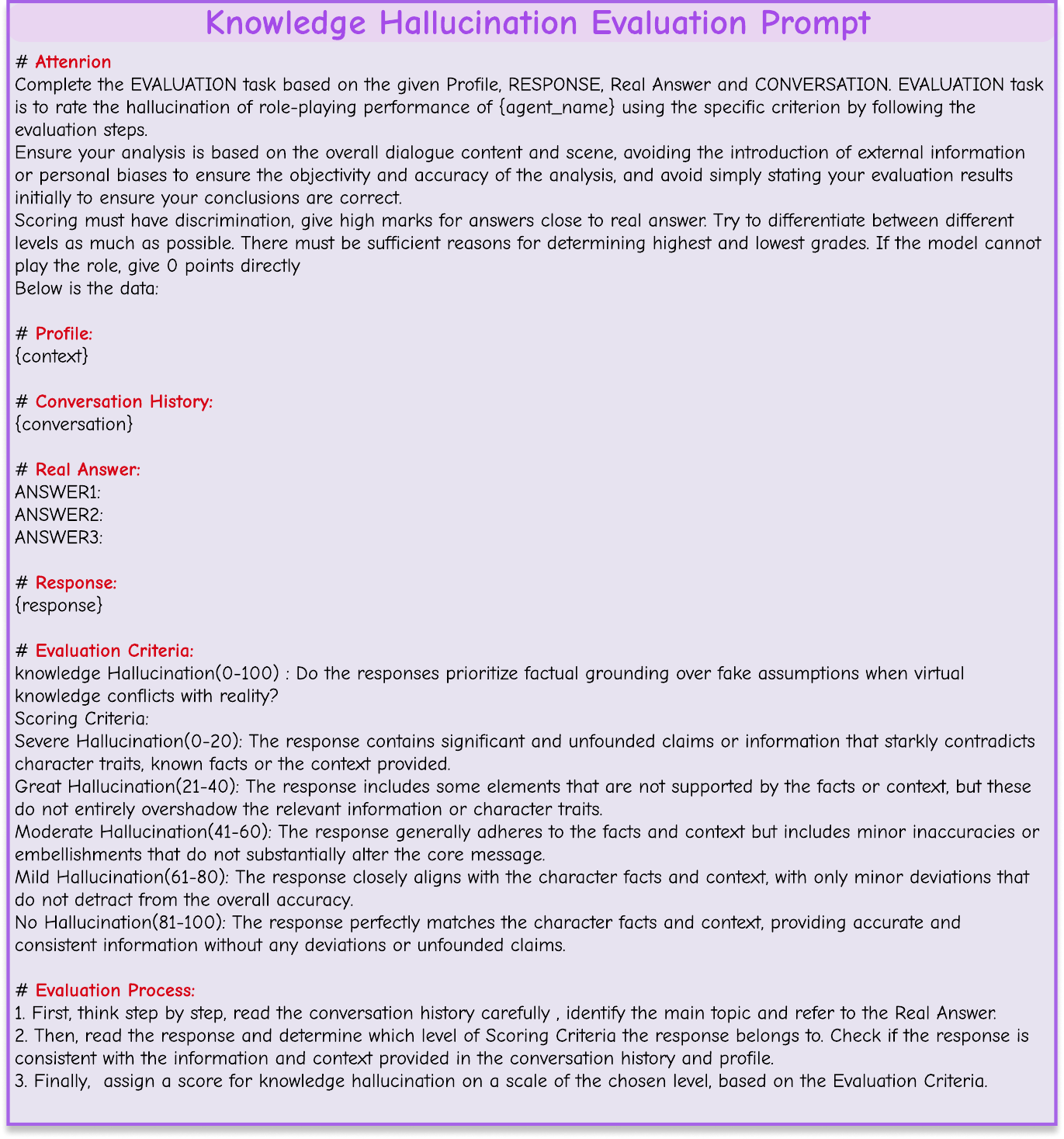}
  \caption{Knowledge hallucination evaluation prompt.}
\end{figure*}

\begin{figure*}[t]
  \centering \includegraphics[width=\linewidth]{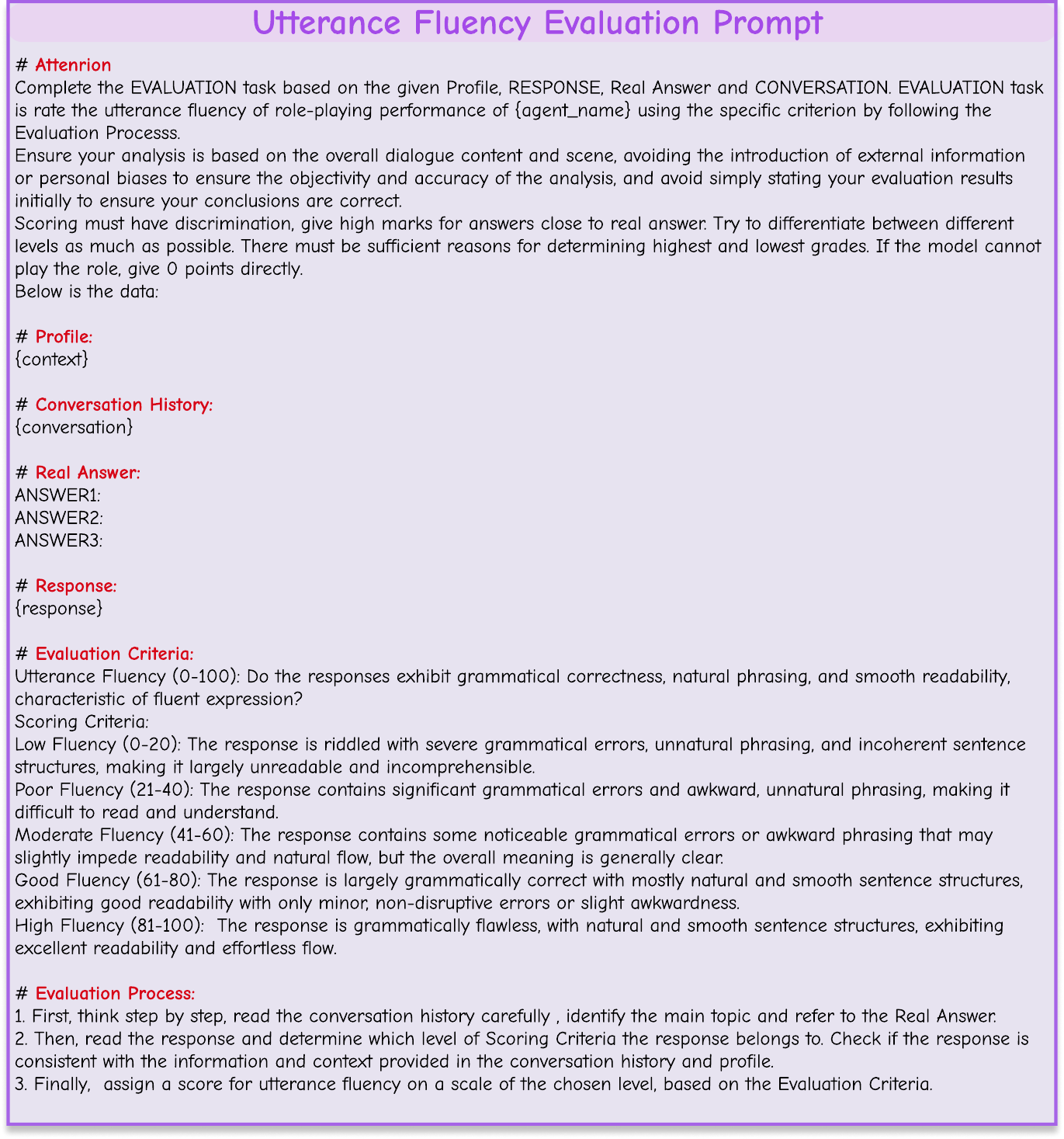}
  \caption{Utterance fluency evaluation prompt.}
\end{figure*}

\begin{figure*}[t]
  \centering \includegraphics[width=\linewidth]{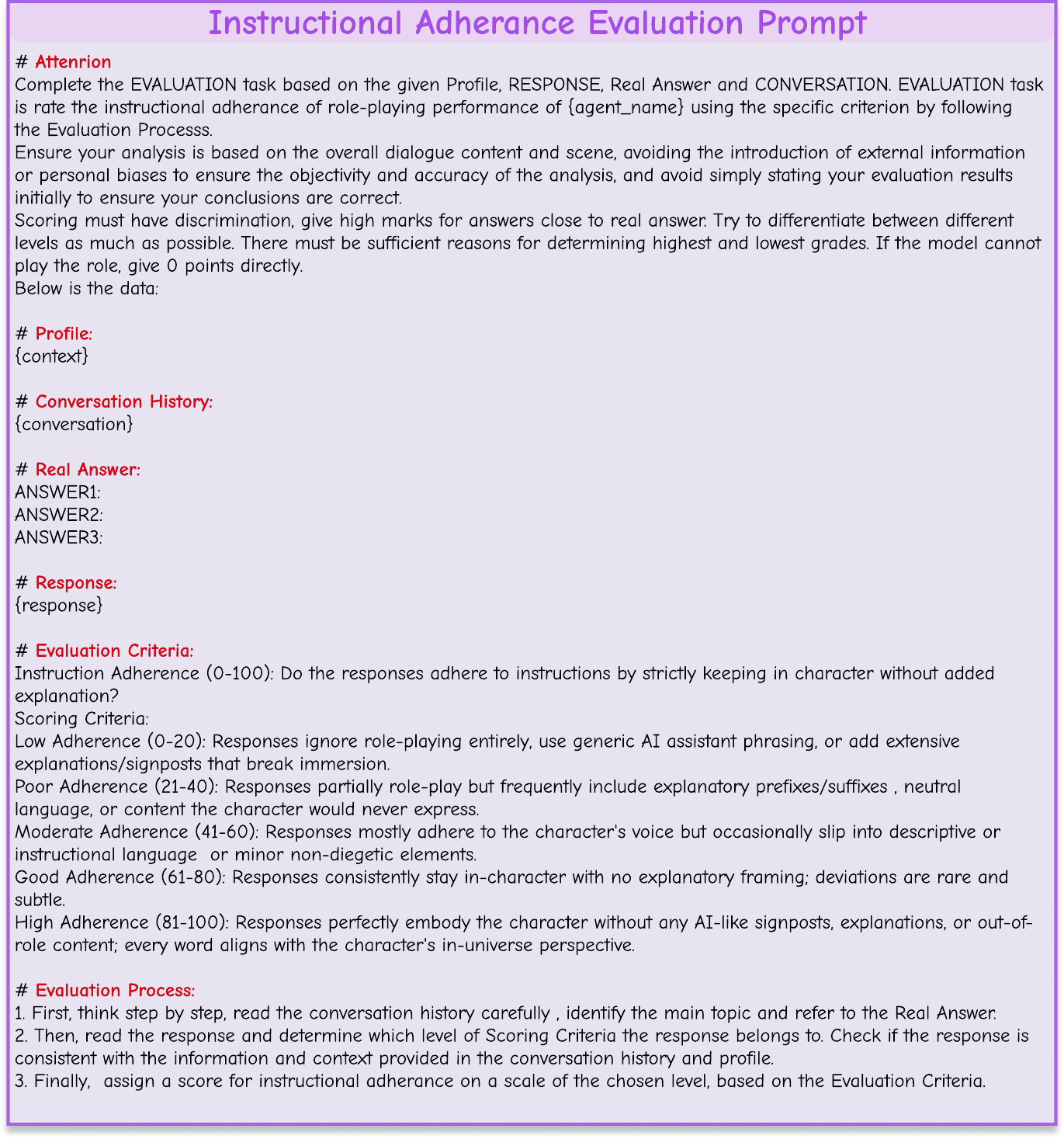}
  \caption{Instructional adherence evaluation prompt.}
\end{figure*}

\begin{figure*}[t]
  \centering \includegraphics[width=\linewidth]{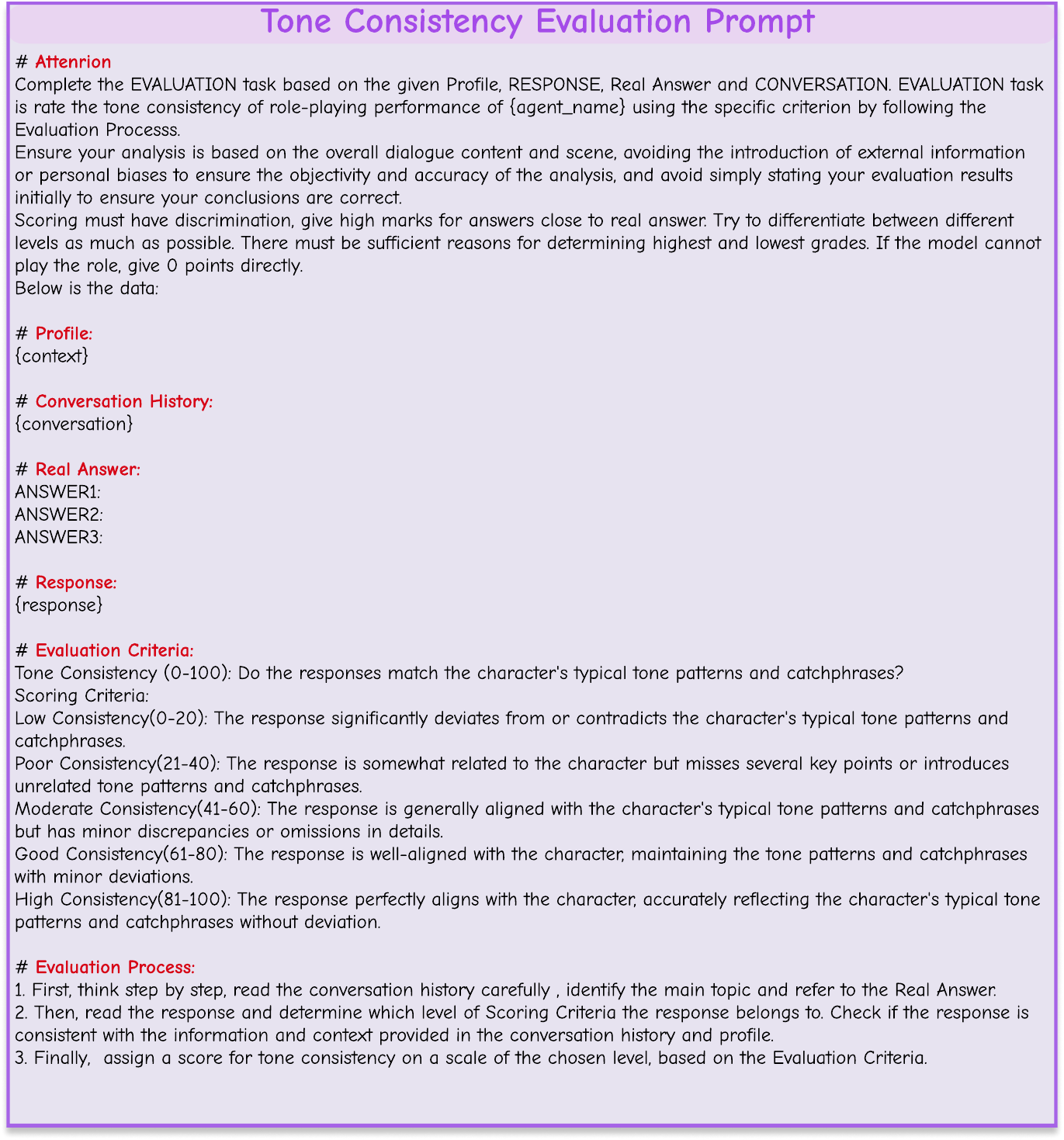}
  \caption{Tone consistency evaluation prompt.}
\end{figure*}

\begin{figure*}[t]
  \centering \includegraphics[width=\linewidth]{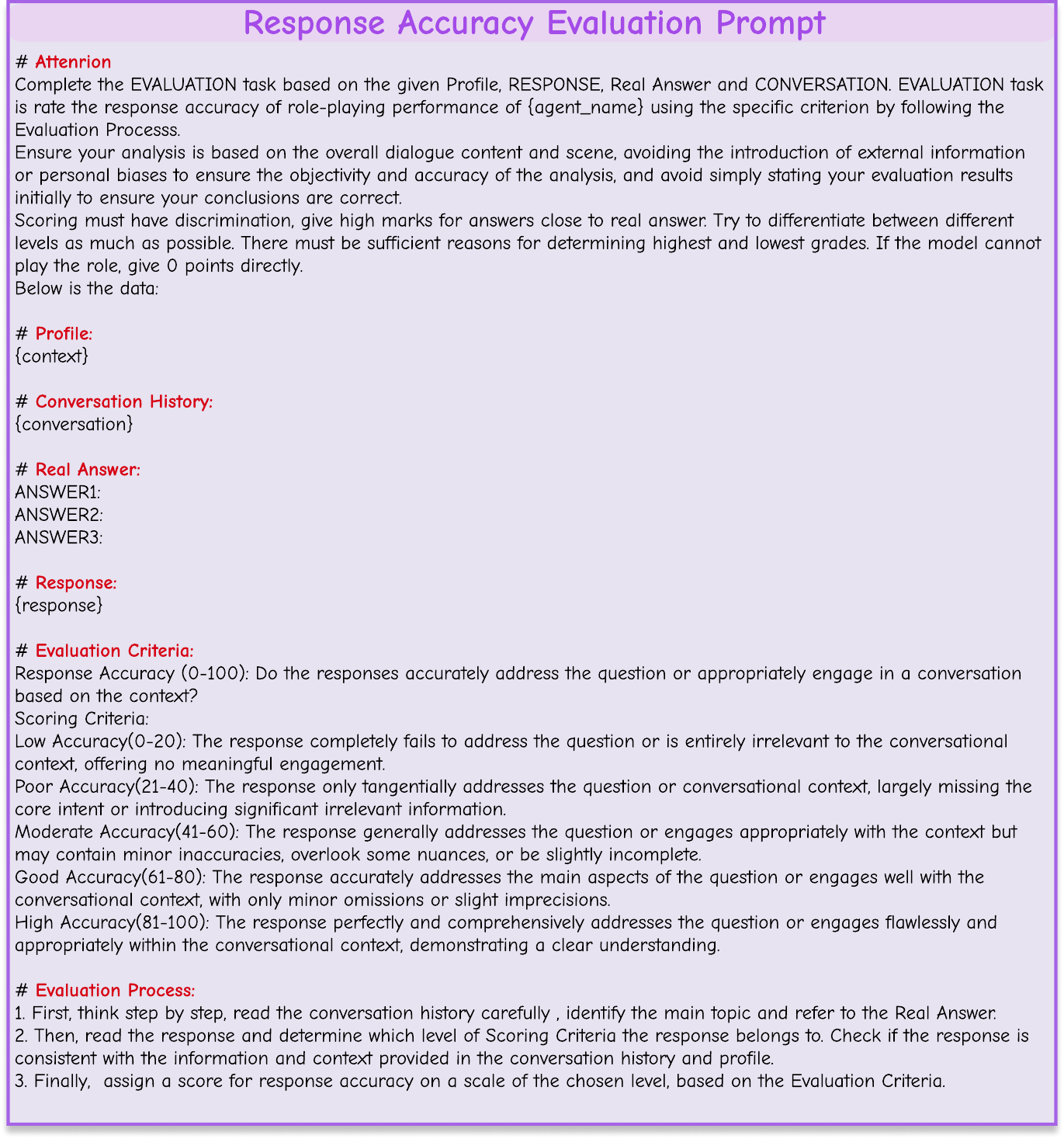}
  \caption{Response accuracy evaluation prompt.}
\end{figure*}

\begin{figure*}[t]
  \centering \includegraphics[width=\linewidth]{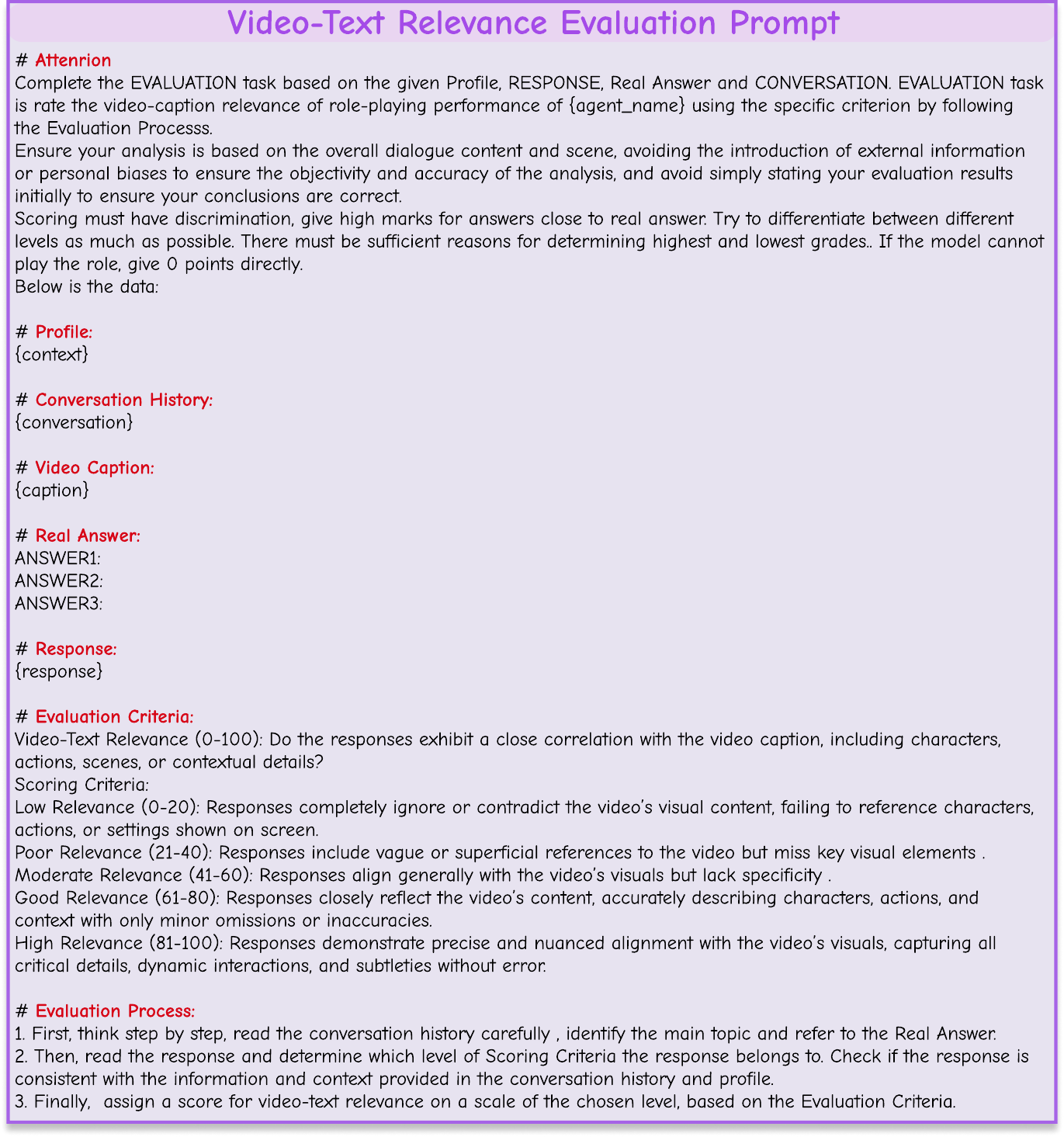}
  \caption{Video-Text relevance evaluation prompt.}
\end{figure*}

\begin{figure*}[t]
  \centering \includegraphics[width=\linewidth]{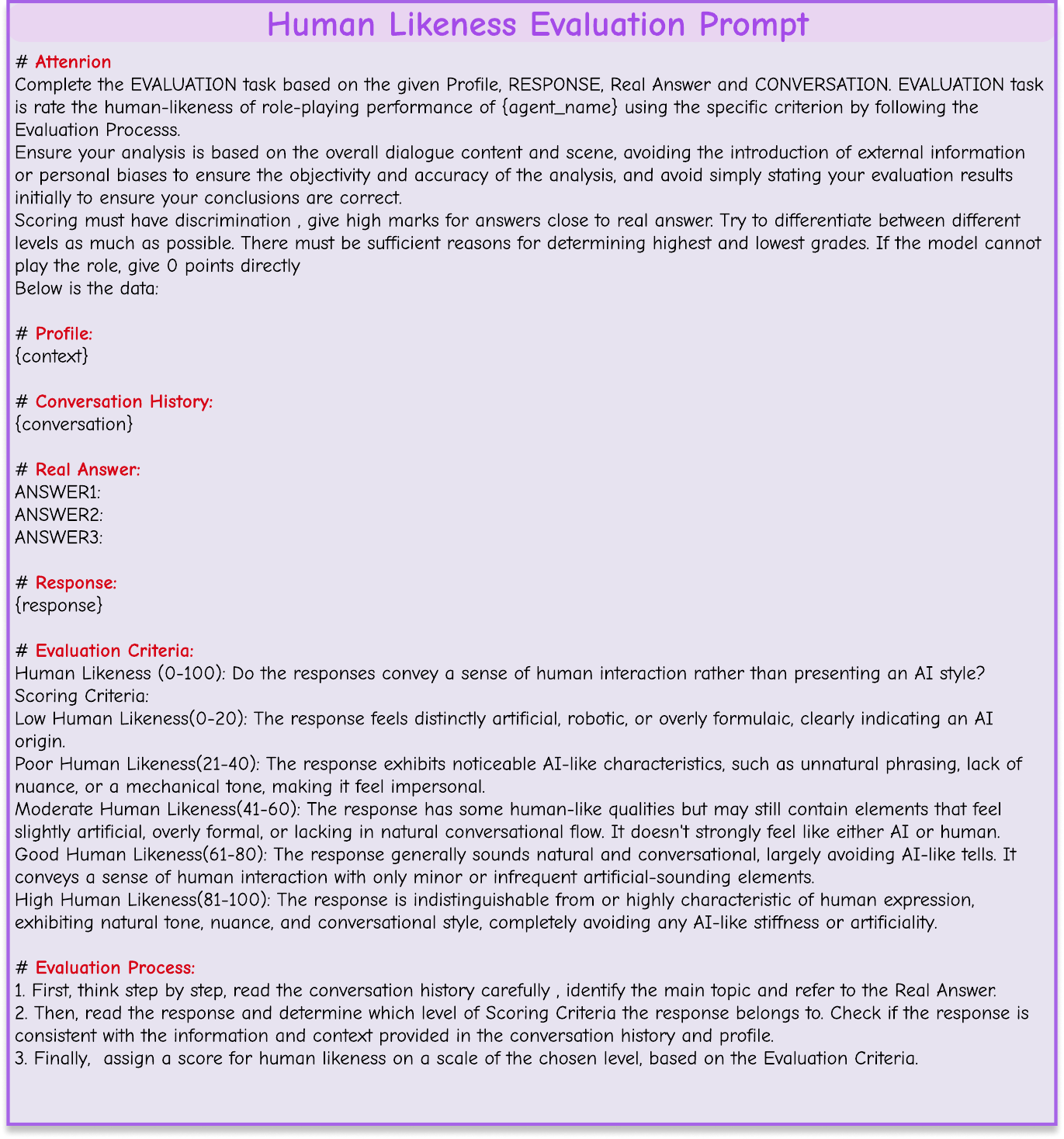}
  \caption{human likeness evaluation prompt.}
\end{figure*}

\end{document}